\documentclass[useAMS,usenatbib, usegraphicx]{mn2e}

\usepackage{amsmath, amssymb}
\usepackage{array}

\begin{document}

\newcommand{\lsim}{\,\raise 0.4ex\hbox{$<$}\kern -0.8em\lower 0.62ex\hbox{$\sim$}\,}
\newcommand{\gsim}{\,\raise 0.4ex\hbox{$>$}\kern -0.7em\lower 0.62ex\hbox{$\sim$}\,}

\title[Principal component analysis of the cosmic microwave background]{Measuring the
  primordial power spectrum: Principal component analysis of the cosmic
  microwave background}

\author[Samuel Leach]{Samuel Leach$^{1,2}$ \thanks{Email address: {\tt leach@sissa.it}}
\\
$^1$SISSA-ISAS, Astrophysics Sector, Via Beirut 4, 34014 Trieste, Italy.\\
$^2$D\'epartement de Physique Th\'eorique, Universit\'e de Gen\`eve,
24 quai Ernest Ansermet, CH-1211 Gen\`eve 4, Switzerland.}

\maketitle

\begin{abstract}
We implement and investigate a method for measuring departures from
scale-invariance, both scale-dependent as well as scale-free, 
in the primordial power spectrum of density perturbations using 
cosmic microwave background (CMB) $C_{\ell}$ data and a principal component
analysis (PCA) technique. The primordial power spectrum is decomposed
into a dominant scale-invariant Gaussian adiabatic component plus a series of
orthonormal modes whose detailed form only depends the noise model for a particular CMB
experiment. However, in general these modes are localised across
wavenumbers with $0.01 < k < 0.2\,{\rm Mpc}^{-1}$, displaying
rapid oscillations on scales corresponding the acoustic peaks where
the sensitivity to primordial power spectrum is greatest. The performance
of this method is assessed using simulated data for the \emph{Planck}
satellite, and the full cosmological plus power spectrum parameter
space is integrated out using Markov Chain Monte Carlo. As a proof of
concept we apply this data compression technique to the current CMB data
from \emph{WMAP}, ACBAR, CBI, VSA and Boomerang. We find no evidence for the breaking of
scale-invariance from measurements of four PCA mode amplitudes, 
which is translated to a constraint on the scalar spectral index  $n_{{\rm
    S}}(k_0=0.04\,$Mpc$^{-1})=0.94\pm0.04$ in accordance with \emph{WMAP} studies.
\end{abstract}



\section{Introduction}
Observations of the cosmic microwave background (CMB) anisotropies are
presenting a fascinating opportunity for discerning between our models
for the origin of structure in the universe in great detail. Indeed
the most recent observations of the CMB from the Wilkinson Microwave
Anisotropy Probe (\emph{WMAP}) have vindicated a basic picture for the primordial
perturbations which are nearly scale-invariant, Gaussian and adiabatic
in nature, and which are dominated by a passive and
growing-mode. This represents enormous progress by instrumentalists in the thirty years
since Zel'dovich and Novikov lamented in their 1975 monologue over the observational
prospects for measuring the CMB anisotropies: 
`\emph{Given all the difficulties, it is not clear that we will ever
  successfully investigate the nature of the initial perturbations
  using the concept of [Sakharov] modulation [of the acoustic
  peaks]\,}' (Zel'dovich \& Novikov 1975). 

At this time, therefore, there is an overall consistency between
observations (Peiris et~al 2003; Barger, Lee \& Marfatia 2003; Leach
\& Liddle 2003) and the inflationary paradigm which is well-known to
contain a mechanism for
generating large-scale perturbations of this type (see Liddle \& Lyth
2000; Dodelson 2003). In the near future though, most progress in our
understanding of the origin of structure is likely to come from
empirical studies of the primordial perturbations where one of the known
ingredients of the standard Gaussian adiabatic model is relaxed to a
more general form. Indeed, this has been the spirit in which many
authors have proceeded. In particular there has been a strong interest in
measuring the shape of the primordial power spectrum, given the
prospect of a factor twenty or so increase in the data to this sector
of cosmology in the near future, coming from ground-based, balloon-borne and
satellite experiments. 

Model-independent methods for reconstructing the primordial power
spectrum are being investigated where one only relies on the broad
assumption that the overall picture of Gaussian adiabatic
perturbations is correct. The available data are then confronted  a
more general primordial power spectrum sector, and the full parameter space is
integrated out in a medium size computation. Many such power spectrum
parametrisations exist and these include bandpowers (Wang, Spergel \&
Strauss 1999; Bridle et~al 2003; Hannestad 2004), 
band-colours (Bridle et~al 2003), wavelet bandpowers (Mukherjee \&
Wang 2003a,c), orthogonal wavelets (Mukherjee \& Wang 2003b). The
specific choices to be made such as the number and the location of the
bandpowers  will require a certain amount of optimisation. However,
these promising methods are known to perform well on both real and simulated
data without degrading too far the expected constraints on the
remaining cosmological parameters (Bond et~al 2004; Mukherjee \& Wang
2005).

One can also apply inverse methods in order to reconstruct
the primordial power spectrum, since the problem at hand is akin to
deconvolution. Many methods have been investigated and these
include semi-analytic iterative methods (Kogo et~al 2005), the
Richardon--Lucy deconvolution algorithm (Shafieloo \& Souradeep 2004),
regularised least-squares (Tegmark \& Zaldarriaga 2002;
Tocchini-Valentini, Douspis \& Silk 2005). While these strategies may provide a
reasonable glimpse of the form of the primordial power spectrum at a lower
computational cost, they typically suffer a weakness that the
cosmological parameters must be fixed to some representative values
and are not integrated out. In addition there is usually a smoothing
step involved either in the data or the deconvolved power spectrum
requiring a careful treatment.

There is a data compression strategy which, although it is most
similar in spirit to the model-independent methods described above,
corresponds to asking a a slightly different question than whether we
can reconstruct or deconvolve the primordial power spectrum. Although the
question we refer to has been in the air and in the minds of many
people for years, and is partially addressed
by any CMB analysis that constrains the power-law slope of the primordial
power spectrum, it is worth stating it here explicitly: \emph{Are
  scale-invariant adiabatic  perturbations an ingredient of our
  cosmology and how can we best measure any departures from
  scale-invariance?} This question is important because its eventual answer
will represent the next step in our attempts to model and understand the
underlying mechanism responsible  generating the primordial
perturbations. We will demonstrate in this paper
that principal component analysis is very well suited for this purpose.
Briefly summarised, the trick is to choose a complete orthonormal
power spectrum basis which also reflects our expectation of
where the departures from scale-invariance are likely to be best probed
by the data, as has been repeatedly emphasised by Hu and collaborators
(Hu \& Okamoto 2004; Kadota et~al 2005). The full cosmological plus
power spectrum parameter space can be integrated out in a medium to
large scale computation, and theoretical predictions for the power
spectrum can be easily projected on onto the same power spectrum basis
to make the comparison with observations.    

The outline of this paper is to describe the principal component
analysis formalism, providing a commentary of the relevant
implementation details in \S \ref{sec:Method}; in
\S \ref{sec:planck} we test the method with simulated \emph{Planck}
data using three primordial power spectra which are respectively
scale-invariant, scale-free, and broken scale-invariant; in
\S \ref{sec:application} we apply the method to the \emph{WMAP} data
before concluding \S \ref{sec:conclusions}.  

\section{Principal component analysis formalism}
\label{sec:Method}
In this paper we implement and investigate the principal component
analysis (hereafter PCA) method detailed and described by Hu and
Okamoto~(2004) (hereafter HO04) which should be considered a companion
paper. PCA has also been applied or discussed in countless other contexts in
which data volumes have already or will soon be seeing sharp
increases, for instance in galaxy-galaxy power spectrum estimation methods
(Hamilton and Tegmark 2000), reionization history reconstruction (Hu
and Holder 2003), dark energy reconstruction (Huterer \& Starkman
2003) and most recently in the context of reconstructing
the inflation potential (Kadota et~al 2005). It can be thought of
simply as a change of parameter basis, where the rotation is
determined by properties of the observed or expected signal and
noise. At the same time it is also a very useful lossless data
compression technique. 

The basic set-up in the context of the CMB is not at all unfamiliar to
astrophysics, that of a deconvolution problem
\begin{eqnarray}
C_{\ell}^{XX'}= \frac{2 \pi}{\ell(\ell+1)} 
\int d \ln k\,{\mathcal P}\left(k\right) 
T^{X}_\ell\left(k;\{\omega_i\}\right) T^{X'}_\ell\left(k;\{\omega_i\}\right),
\end{eqnarray}
where $X=T,E$ and the dependence of the CMB transfer functions
$T^{X}_\ell\left(k\right)$ on the cosmological parameters 
$\{\omega_i\}$ has been written explicitly
in order to show the added complication over and above an ordinary
deconvolution problem of this type. Interestingly, there is a
satisfactory solution to the problem of extracting the primordial power
spectrum ${\mathcal P}(k)$, described in HO04, which involves
exploiting what we know about the expected noise on $C_{\ell}$ and our
precise and accurate knowledge of the CMB transfer function physics (Seljak
et~al 2003). Here we present the relevant equations from HO04.

The response of the
$C_{\ell}$ with respect to some primordial power spectrum parameters
$\{p_i\}$ can be investigated via a mode counting approach by
constructing the Fisher information matrix 
\begin{eqnarray}
F_{ij} = 
\sum_{\ell=2}^{\ell_{\rm max}}
\frac{2 \ell +1}{2} 
{\rm Tr} [{\bf D}_{\ell i} 
{\bf C}_{\ell}^{-1}
{\bf D}_{j \ell} {\bf C}_{\ell}^{-1}]\,,
\label{eqn:fisher}
\end{eqnarray}
which has been written using a matrix notation, where
\begin{eqnarray}
({\bf D}_{\ell i})_{XX'} =  D_{\ell i}^{XX'} 
\equiv \frac{\partial C_\ell^{XX'}}{ \partial p_i},
\end{eqnarray}
and where
\begin{eqnarray}
D_{\ell i}^{XX'} &=& \frac{\partial C_\ell^{XX'}}{ \partial p_i}\big|_{\rm fid} \label{eqn:derivdef}\\
&=& \frac{2 \pi}{ \ell(\ell+1)} \int d\ln k\,{\mathcal P}_0
T^{X}_\ell(k) T^{X'}_\ell(k)\,  W_i(\ln k)\,.
\nonumber
\end{eqnarray}
We can take our power spectrum test function $W_i$ to be
the triangle window
\begin{eqnarray}
W_i(\ln k) =  {\rm max}\left[1- \left|\frac{\ln k -\ln k_i}{ \Delta \ln k}\right|,0\right] \,.
\label{eqn:window}
\end{eqnarray}
In this work we have used a discretisation $\Delta \ln k = 0.00875$
spanning a range of scales that traverses the acoustic peaks
from $0.004< k < 0.2\, {\rm Mpc}^{-1}$. It is worth noting at this stage
that this range need not include the largest scales
responsible for the Sachs--Wolfe effect: the Fisher information
on these scales tends to zero, and so it proves convenient to truncate
these scales in order to later on invert the Fisher information matrix without
numerical difficulties. The calculation of the power spectrum transfer
functions $D_{\ell i}^{XX'}$ is
achieved by making minor modifications to the {\sc CAMB} CMB
anisotropies code (Lewis, Challinor \& Lasenby 2000) (based on {\sc CMBFAST}
(Seljak \& Zaldarriaga 1996)), rather than
using a full Boltzmann hierarchy code used in HO04. {\sc CAMB} is run
at slightly higher accuracy where we have increased by a factor four
both the number of source and integration $k$ modes, and have
calculated $D_{\ell i}^{XX'}$ at every $\ell$, rather than the usual
splining method with $\Delta \ell \sim 50$, in order to capture the
high frequency information. 

\begin{figure}
\centering
\includegraphics[width=\linewidth]{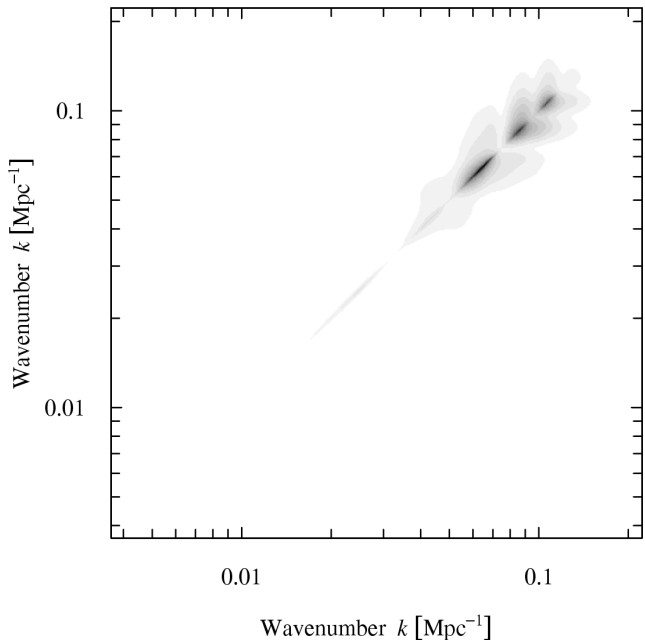}
\caption{Illustrating $F_{ij}$, given by equation (\ref{eqn:fisher}), for the
  \emph{Planck} satellite, which displays a band-diagonal structure with
  peaks in sensitivity corresponding to the temperature acoustic
  peaks. Here the discretisation is $\Delta\ln k=0.00875$. The
  bandwidth of the Fisher matrix, $\delta \ln k\sim 0.05$,
  determines the maximum achievable resolution for the recovery of
  the primordial power spectrum.  
}\label{fig:FisherMatrix}
\end{figure}

The choice of fiducial cosmological parameters is given by a baryon
density $\Omega_{{\rm B}}h^2=0.024$, cold dark matter density
$\Omega_{{\rm     D}}h^2=0.121$, present Hubble rate $H_0\, [{\rm
  km}\, {\rm s}^{-1}{\rm Mpc}^{-1}] =72$, optical depth to last
scattering $\tau=0.17$, and a curvature perturbation amplitude
${\mathcal P}_0=23\times10^{-10}$. We assume a spatially flat
cosmology and ignore the effect of lensing. The latter will be
important to take into account in a more thorough analysis in order
avoid biasing of the recovered cosmological parameters (Hu \& Okamoto
2004; Lewis 2005). 

In Fig.~\ref{fig:FisherMatrix} we illustrate the Fisher information
matrix given by equation (\ref{eqn:fisher}) which shows a band-diagonal
structure with peaks of sensitivity to the primordial power spectrum on
scales corresponding to the acoustic peaks; the sensitivity drops
again on scales corresponding 
to the acoustic troughs, which can be remedied by
information coming from the phase-shifted polarization peaks. Of
course the sensitivity tends to zero on large scales due to a lack of
modes to observe, and on small scales due to Silk damping and beam
smoothing, since the $C_{\ell}$ of
equation (\ref{eqn:fisher}) is replaced by the total signal plus a Gaussian
white noise model adjusted for a given experiment
\begin{eqnarray}
\label{eqn:gaussiannoisemodel}
C_\ell^{TT}\big|_{\rm noise} &=& 
\sigma_{{\rm noise}}^2 e^{\ell(\ell+1) \theta^2/8\ln 2}\,,\nonumber \\
C_\ell^{EE}\big|_{\rm noise} &=& 2\times\sigma_{{\rm noise}}^2
e^{\ell(\ell+1) \theta^2/8\ln 2}\,,\\
C_\ell^{TE}\big|_{\rm noise} &=& 0,\nonumber
\end{eqnarray}
where $\sigma_{{\rm noise}}^2$ is the noise variance in
$(\mu$K-rad$)^2$ and $\theta$ is the FWHM of a Gaussian beam in
radians. The noise model should be considered an important input to
the analysis since it determines the range of scales that will be
probed; it is an additional ingredient
compared to the majority of analyses of the $C_{\ell}$
data.
We use here a noise model for \emph{Planck} with
$\sigma_{{\rm     noise}}^2=3\times10^{-4}(\mu$K-rad$)^2$ and $\theta
= 7'$, and a noise model for \emph{WMAP} with 
$\sigma_{{\rm noise}}^2=8.4\times10^{-3}(\mu$K-rad$)^2$ and $\theta =
13'$. In a realistic analysis the observed signal plus noise spectrum
will be more appropriate.    

\begin{figure}
\centering
\includegraphics[width=\linewidth]{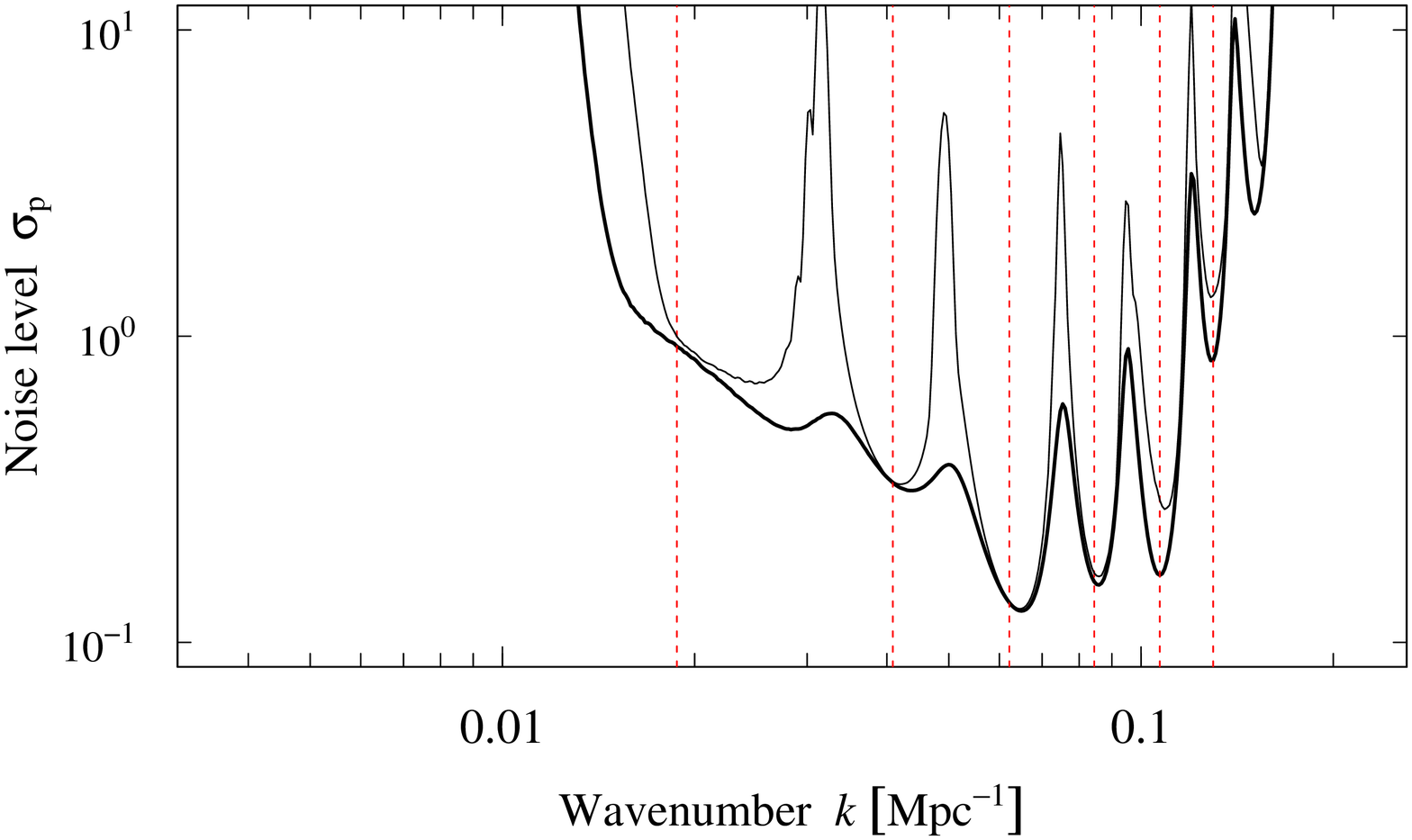}
\caption{Illustrating \emph{Planck}'s window of sensitivity to the
  primordial power spectrum with and without polarisation (upper
  curve). Here $\sigma_p$ gives the approximate 1$\sigma$ error on
  measurements of the primordial power spectrum using bandpowers
  with $\delta \ln k\sim0.02\rightarrow 0.05$. The vertical lines indicate the
  position of the temperature acoustic peaks.
  The cosmological parameters have been fixed, so some degrading
  of the sensitivity is expected.
}\label{fig:planckpolnopoldiag}
\end{figure}

As usual the Fisher information matrix can be inverted to obtain a
covariance matrix $C_{ij}$ whose diagonal components provide a useful
estimate, the Cramer--Rao bound, of the expected variance of the
parameters $p_i$ with 
\begin{eqnarray}
\label{eqn:1sigma}
\sigma^2(p_i)= C_{ii}\approx (F^{-1})_{ii}.
\end{eqnarray}
In Fig.~\ref{fig:planckpolnopoldiag} we plot this window of
sensitivity to the primordial power spectrum (on a scale $\delta \ln
k\sim0.05$ set by the Fisher matrix bandwidth) for the \emph{Planck}
satellite, which can be seen to encompass the entire acoustic peak
region. As noted in HO04, outside this range of scales, and in
particular on large scales, we can only hope to recover wide-band
($\delta \ln k \gg 0.05$) averages of the primordial power spectrum at high
accuracy.

\begin{figure}
\centering
\includegraphics[width=\linewidth]{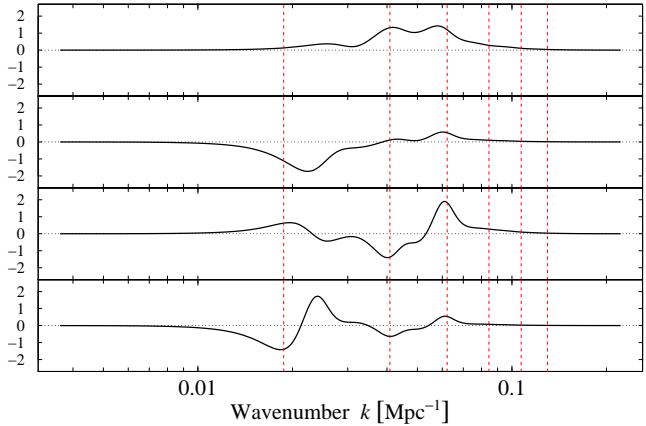}
\caption{Illustrating PCA modes 1--4 which have been generated assuming
  the \emph{WMAP} noise model. The vertical lines indicate the
  position of the temperature acoustic peaks.
}\label{fig:modes1to4}
\end{figure}

\begin{figure}
\centering
\includegraphics[width=\linewidth]{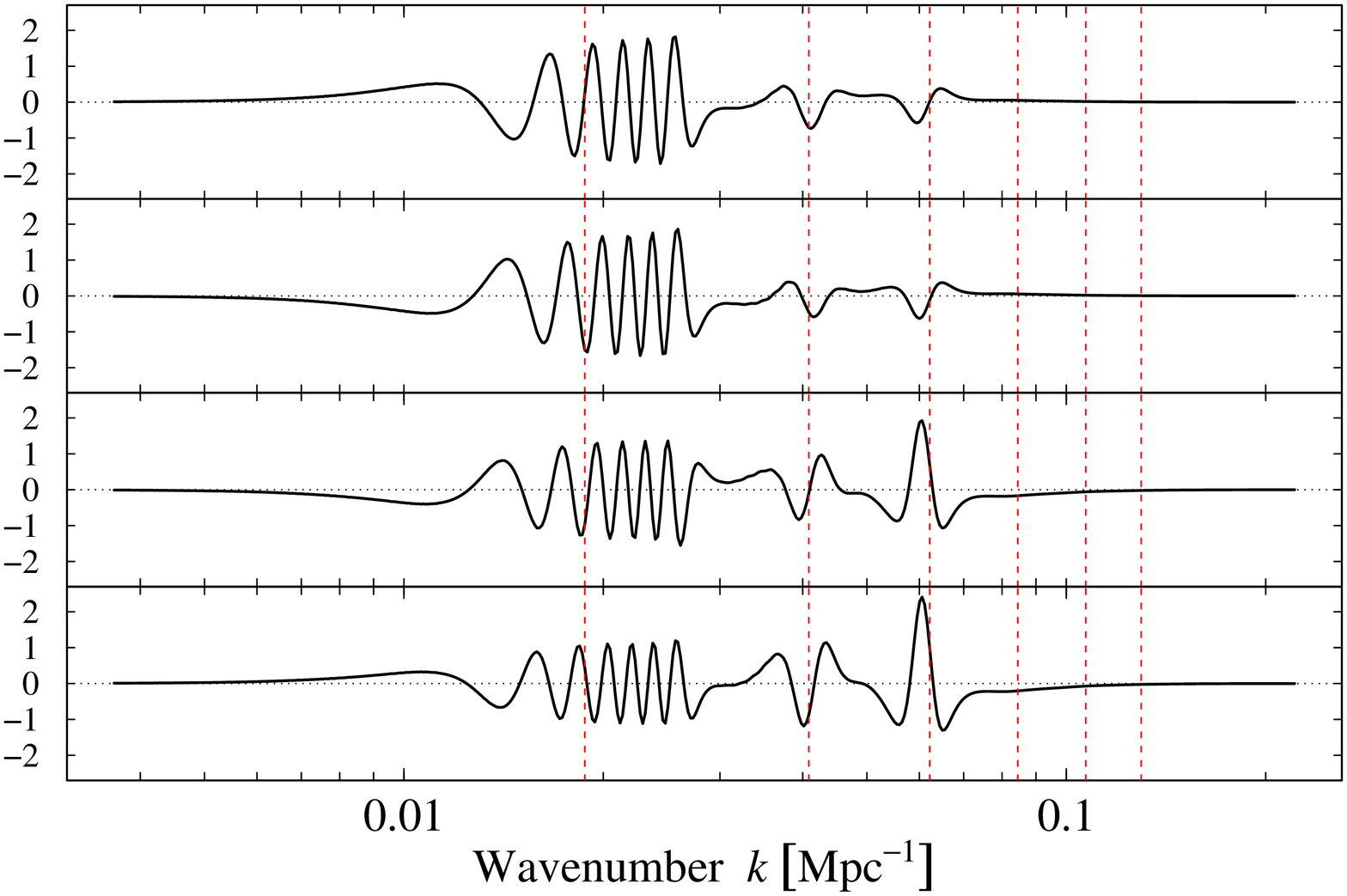}
\caption{Illustrating PCA modes 17--20, as in
  Fig.~\ref{fig:modes1to4}. The oscillations are strongest in the
  vicinity of the acoustic peaks where the sensitivity to the primordial
  power spectrum is greatest.}\label{fig:modes17to20}
\end{figure}

The PCA basis $\{m_i\}$ is simply a linear combination of the power spectrum
spike basis $\{p_i\}$
\begin{eqnarray}
m_a = (\Delta \ln k)^{1/2} \sum_i S_{ia} p_i
\end{eqnarray}
where the $S_{ia}$ are the orthonormal eigenvectors of the covariance
matrix. We can then work with a set of normalised principal components
${\mathcal S}_{ia}=S_{ia}/\sqrt{\Delta \ln k}$ (hereafter the PCA modes)
which will have unit variance when integrated over $\ln k$. In
Figs.~\ref{fig:modes1to4} and \ref{fig:modes17to20} 
we plot examples of the PCA modes with mode number from 1--4 and 17--20
respectively, generated using the \emph{WMAP} noise model. 
The oscillations in the PCA modes become increasingly
rapid at scales corresponding to the acoustic peaks where sensitivity
to the primordial power spectrum is greatest, that is until we hit the
numerical resolution. At this point the PCA modes branch into two
wavepacket-like solutions travelling towards large and
small scales, similar to the behaviour noted by Hamilton and Tegmark
(2000), although this need not worry us.
Note also that the PCA modes are invariant under changes in
the discretisation scale $\Delta \ln k$. However, we found that in
order to obtain sensible estimates of the eigenvalues (projected
errors) of the PCA modes themselves, the Fisher matrix should be
discretised on a scale that renders it roughly diagonal, instead of
band-diagonal.

The PCA modes can be straightforwardly integrated into the publicly available
Markov Chain Monte Carlo (MCMC) package {\sc CosmoMC}\footnote{{\tt
    http://cosmologist.info/cosmomc/}} (Lewis \& Bridle 2002, February
2005 version) in order to explore the full cosmological plus power
spectrum posterior parameter space. Specifically, we use the following
power spectrum ansatz
\begin{eqnarray}
\label{eqn:PofkParametrisation}
\frac{{\mathcal P}(k)}{{\mathcal P}_0}= m_0+ \sum_{a=1}^{a_{{\rm
        max}}}m_a {\mathcal S}_a(k),
\end{eqnarray}
where we take ${\mathcal P}_0=23\times 10^{-10}$, which should be
calibrated from observations. Clearly if the
underlying primordial power spectrum is close to scale-invariant then
equation (\ref{eqn:PofkParametrisation}) admits a solution
\begin{eqnarray}
m_a = 0,\,\, \forall a \Leftrightarrow \mbox{\rm Scale-invariance}.
\end{eqnarray}
More generally equation (\ref{eqn:PofkParametrisation}) is strongly suggestive of
a general linear orthonormal model plus a noise term (see for instance
Bretthorst 1988). In this way we are attempting to measure the
spectrum of departures from scale-invariance which we call $\Delta
{\mathcal P}/{\mathcal P_0}$ and which is given by the second term in
equation (\ref{eqn:PofkParametrisation}); in this context the dominant
scale-invariant component $m_0$ is a Gaussian noise term.

Concerning the numerical implementation of the power spectrum
equation (\ref{eqn:PofkParametrisation}), we simply perform a linear spline
in $\ln k$ over the discrete PCA modes ${\mathcal S}_{ia}$, which are
added together before the final convolution with CMB transfer functions
to obtain the $C_{{\ell}}$; outside the PCA mode $k$-range the
second term of equation (\ref{eqn:PofkParametrisation}) is dropped. We
checked that the default $k$-source and $k$-integration settings for
{\sc CAMB}  modified to calculate $C_{{\ell}}$ at $\Delta \ell = 3$ is
accurate enough handle around the first forty modes of our current
implementation; at this stage this is more than enough since we will
only attempt to perform the MCMC with a maximum of sixteen PCA modes.  

Having obtained measurements of the PCA mode amplitudes from the MCMC,
it is then straightforward to project any power spectrum model, for
instance a power-law spectrum, onto the
PCA modes via
\begin{eqnarray}
\label{eq:powerproject}
m_a &=& \int d \ln k \,{\mathcal S}_a(k) \frac{\Delta {\mathcal
    P}}{{\mathcal P_0}}(k),\nonumber\\
    &= &\Delta \ln k \sum_i {\mathcal S}_a(k_i)\left[\left( \frac{k_i}{k_0}\right)^{n_{{\rm
      S}}-1} -1\right],
\end{eqnarray}
in order to make the comparison with observations.

We can easily make an empirical estimate of the total signal to noise
of the measured departures from scale-invariance
\begin{eqnarray}
  \frac{\rm S}{\rm N}=\left[\sum_{a=1}^{a_{{\rm max}}}
    \frac{\left<m_a\right>^2}{\sigma^2_{m_a}} \right]^{1/2}, 
\end{eqnarray}
where $\left<m_a\right>$ and $\sigma^2_{m_a}$ are the mean and variance of the
individual mode amplitudes obtained from the MCMC.
As noted by Kadota et~al (2005), the PCA modes can be safely truncated as
soon as ${\rm S}/{\rm N}$ saturates, assuming that the underlying
primordial power spectrum is a reasonably smooth
function. Incidentally, the total ${\rm S}/{\rm N}$ represents a useful
figure of merit for optimising future CMB experiments to measure
the primordial power spectrum sector. Other measures such as ``risk''
(Huterer \& Starkman 2003) and Bayesian evidence (see for example
MacKay 2003) could be used to provide a rationale for
truncating the PCA mode amplitudes even further, given a
power spectrum model of interest.

In the case that the recovered PCA mode amplitudes encode some complex
information which can not be easily understood in the framework of
power-law spectra, then it would be useful to obtain an estimate 
of $\Delta {\mathcal P}/{P_0}$ in $k$-space in order to aid the
process of modelling the power spectrum. Here we use an estimator
\begin{eqnarray}
\label{eq:meandeparture}
\frac{\hat{\Delta {\mathcal P}}(k_i)}{{\mathcal P}_0}=
   \sum_{a=1}^{a_{{\rm max}}}\left<m_{a}\right> {\mathcal S}_{a}(k_i),
\end{eqnarray}
and for the purposes of a comparison with the input spectrum, we
estimate the noise variance via 
\begin{eqnarray}
\label{eq:esterror}
\hat\sigma^2_{\frac{\Delta{\mathcal P}}{{\mathcal P}_0}}(k_i) =
C_{ii}+\sum_{a=1}^{a_{{\rm max}}}{\mathcal S}_{a}^2(k_i)\sigma_{m_a}^2,
\end{eqnarray}
where $C_{ii}$ is the covariance matrix, obtained from
equation (\ref{eqn:1sigma}), accounting for the overall 
uncertainty in the  narrow-band determination of 
${\Delta{\mathcal P}}/{{\mathcal P}_0}$ in regions of
lower sensitivity on large scales, small scales, and in the
temperature acoustic trough regions.

A bandpower representation of the primordial power spectrum
could also obtained from the measured PCA mode amplitudes via a Monte Carlo
procedure; in this case the Fisher information matrix could be used
for guidance when choosing the location and widths of the
bands. Obviously though, no further quantitative information about the
primordial power spectrum can be gleaned in this way.    

\begin{figure}
\centering
\includegraphics[width=\linewidth]{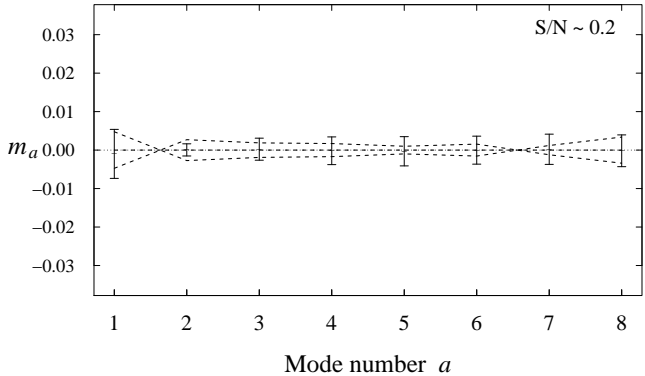}
\caption{Illustrating the recovery of the first eight mode amplitudes
  from simulated \emph{Planck} data with an input scale-invariant
  spectrum. Plotted are the marginalised 1$\sigma$ error bars  obtained from
  MCMC. The models (dashed lines) are
  for power-law spectra with $n_{{\rm S}}(k_0=0.05$ Mpc$^{-1})=\{0.99,1,1.01\}$ (top   to
  bottom, mode 1).  
}\label{fig:PLANCK89_PCAspectrum}
\end{figure}

One final point worth making in this section concerns how one should
deal with the inevitable degeneracies between the effect on the
$C_{\ell}$ due to the cosmological parameters and 
the PCA power spectrum parameters, which will induce undesired
off-diagonal components in the PCA covariance matrix. We sketch here
the solution given in HO04: One must first form the joint
Fisher information matrix,  $F_{\mu\nu}$, for both power spectrum
parameters and cosmological parameters
\begin{eqnarray}
  F_{\mu\nu}=\left[
    \begin{array}{rrr}
    F_{ij} && B \\
    B^{\rm T}&& F_{ab}
\end{array} 
\right],
\end{eqnarray}
where $F_{ab}$ is the usual cosmological parameter Fisher information
matrix (see for example Tegmark, Taylor and Heavens 1997)
and $B$ are the cross terms. After inverting the full $F_{\mu\nu}$
to obtain a new covariance matrix  $C_{\mu\nu}$, one simply
retains the power spectrum parameter subblock $C_{ij}$, whose principal
components will be ``orthogonalized'' to the effect of the
cosmological parameters. In terms of implementation, one can use the
matrix partitioning formulas (see for example Press et~al 1992, \'O
Ruanaidh \& Fitzgerald 1996) to derive a ``degraded'' $F^{\rm deg}_{ij}$
subblock 
\begin{eqnarray}
F^{\rm deg}_{ij}= F_{ij}-B^{\rm T}F_{\rm ab}B.
\label{eqn:degrade}
\end{eqnarray}
We will make use of this in the next section.

\begin{figure}
\centering
\includegraphics[width=\linewidth]{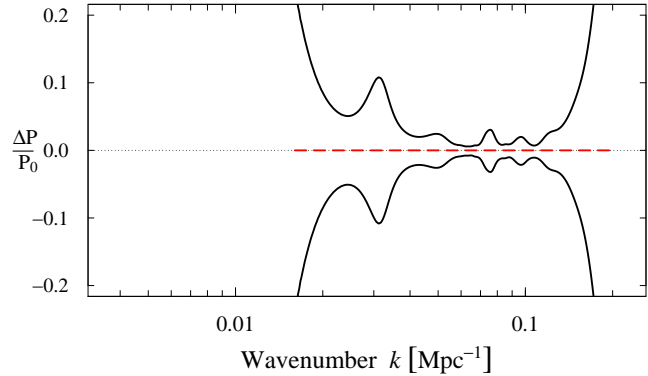}
\caption{Illustrating the estimated departures from scale-invariance in
  $k$-space on a narrow-band scale $\delta \ln k \sim 0.02$ 
  for the case of an input scale-invariant spectrum. The
  solid curves show the estimated 1$\sigma$ error bars,
  given by equation (\ref{eq:esterror}). A scale-invariant spectrum within
  the acoustic peak region is strongly favoured.  
}\label{fig:HZ8}
\end{figure}

\section{Tests with Simulated \emph{Planck} data}
\label{sec:planck}
As a means of gaining experience with the PCA method 
we generate simulated \emph{Planck} data up to an
$\ell_{{\rm max}}=2250$ using the Gaussian white noise model of
equation (\ref{eqn:gaussiannoisemodel}) for a cosmological model with parameters
$\Omega_{{\rm B}}h^2=0.024$, $\Omega_{{\rm D}}h^2=0.121$, $H_0=72$,
$\tau=0.17$, and ${\mathcal P}_0=2.3\times10^{-9}$, which for simplicity are the same
as those used to generate the PCA modes themselves. In a realistic
data analysis scenario, the PCA modes would be generated with
parameters close to the best-fit obtained from a traditional parameter
determination approach. We consider three
cases for the primordial power spectrum  which is taken to
described by a scale-invariant spectrum, a power-law 
spectrum with spectral index $n_{{\rm S}}=0.97$ and pivot scale
$k_0=0.05\,{\rm  Mpc}^{-1}$, and then finally a broken
scale-invariance model with a Gaussian bump in the acoustic peak region
\begin{eqnarray}
\label{eq:bump}
  \frac{{\mathcal P}(k)}{{\mathcal P}_0}=1 +
    0.1\exp\left[-\left(\frac{\ln\left[{k}/{0.08\,{\rm Mpc}^{-1}}\right]}{0.3}\right)^2\right]. 
\end{eqnarray}
We then perform MCMC over the full cosmological plus PCA mode
parameter space using the simulated data up to an $\ell_{\rm
  max}=2000$. We have also varied the number of modes included
in the analysis from zero to sixteen in steps of four in order to
study the effect of truncating the PCA expansion on the recovery
of the cosmological parameters. 

The development of {\sc CosmoMC} (Lewis \& Bridle 2002) has
reached a maturity that is very well suited to an analysis of
this type where the number of power spectrum parameters begins to
dominate over the number of cosmological parameters, but where we
nonetheless expect by construction to obtain a stable multivariate
Gaussian posterior solution. As a result we have taken full advantage of
a conjugate gradients descent  module which estimates the covariance and
location of the posterior peak before the MCMC begins, thus
alleviating the potential challenge  working with so many parameters
while also conserving some computing resources. On this note, the
total number of $C_{\ell}$ likelihood evaluations required in our
tests in the following section rises from around ${\mathcal
  N}_{\mathcal   L}=2\times 10^4 \rightarrow 10^6$ for zero and eight
PCA modes respectively, and then tends to saturate at around this
number. It seems reasonable that the number of likelihood evaluations
ought not to exceed by much $\ell^2$, the total number of modes upon
which the the $C_{\ell}$ spectrum depends.
Moreover the `fast--slow' split between power spectrum and
cosmological parameter likelihood  evaluation speeds, already
implemented in {\sc CosmoMC}, will be of increasing benefit as we
attempt to measure up to perhaps thirty PCA mode amplitudes in the
future (Kadota et~al 2005).

\subsection{The scale-invariant case}

In Fig.~\ref{fig:PLANCK89_PCAspectrum} we illustrate the
recovery of the first eight mode amplitudes for the
$n_{{\rm S}}=1$ case and make comparison for the theoretical
prediction for the mode amplitudes which are obtained by
projecting some representative power-law spectra onto the PCA modes
via equation (\ref{eq:powerproject}); we find that the scale-invariant
solution $m_a=0$ is very well recovered.
Here it is worth mentioning that the Gaussian realisation for the
simulated \emph{Planck} data sets was taken to be the exact $C_{\ell}$ model,
which explains why the recovery of the PCA mode amplitudes shows very
little scatter around $m_a=0$. One can see that the first three PCA
modes provide the bulk of constraining power for smooth power-law spectra
leading to a constraint which will be roughly $n_{\rm S}=1\pm0.01$,
consistent with typical parameter forecasts in the literature.

\begin{figure}
\centering
\includegraphics[width=\linewidth]{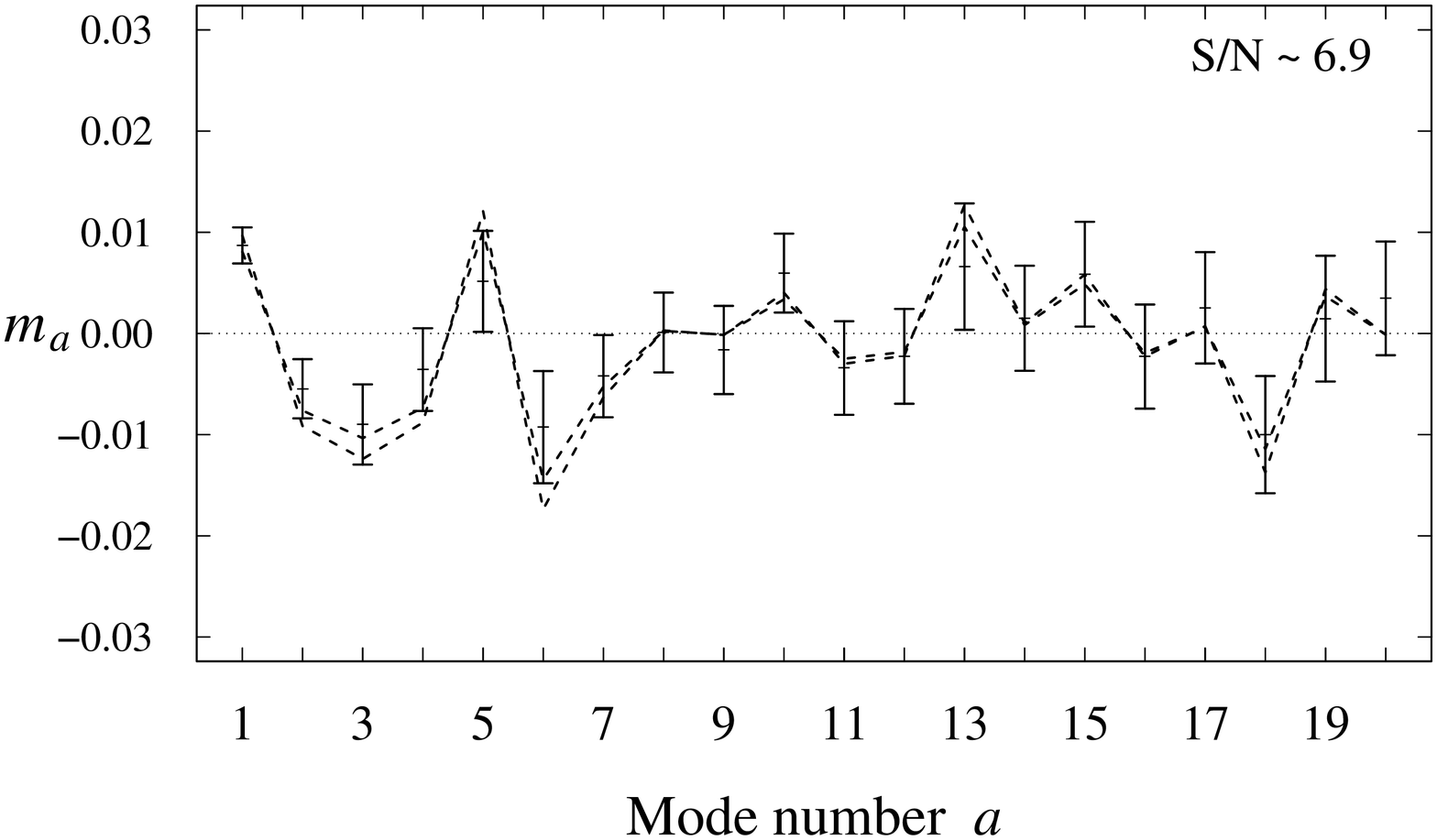}
\caption{Illustrating the recovery of the first ten principal
  component amplitudes from simulated \emph{Planck} data with an input $n_{{\rm
      S}}=0.97$ spectrum, as in Fig.~\ref{fig:PLANCK89_PCAspectrum}.
  The models (dashed lines) correspond to power-law spectra with $n_{{\rm
      S}}(k_0=0.05$ Mpc$^{-1})=\{0.97,0.975\}$ (bottom to top, mode
  3). The compressed CMB data can not be fit by $m_{\rm a}=0$
  and so scale-invariance would be ruled out at high signal to noise.   
}\label{fig:ns097PCASpectrum8}
\end{figure}

We illustrate  an estimate of the departures from scale-invariance 
$\hat{\Delta {\mathcal P}}/{\mathcal P}_0$ in
Fig.~\ref{fig:HZ8}, and the region with the most data weight
can clearly be discerned showing consistency with a
scale-invariant spectrum.
In this case the recovery of the cosmological parameters is also excellent,
and we recovered a stable Gaussian posterior (as a function of the
number of PCA modes)  with
constraints given by $\omega_{\rm B}h^2=0.0240\pm0.0002$, $\omega_{\rm
  D}h^2=0.121\pm0.02$, $H_0=71.9\pm0.7$,
$\tau=0.170\pm0.005$, ${\mathcal P}/{\mathcal P_0}=1.00\pm0.01$ for the
case of using eight PCA modes. Clearly the PCA method works well under
these most idealised of circumstances.

\subsection{The  scale-free $n_{{\rm S}}=0.97$ case}

\begin{figure}
\centering
\includegraphics[width=\linewidth]{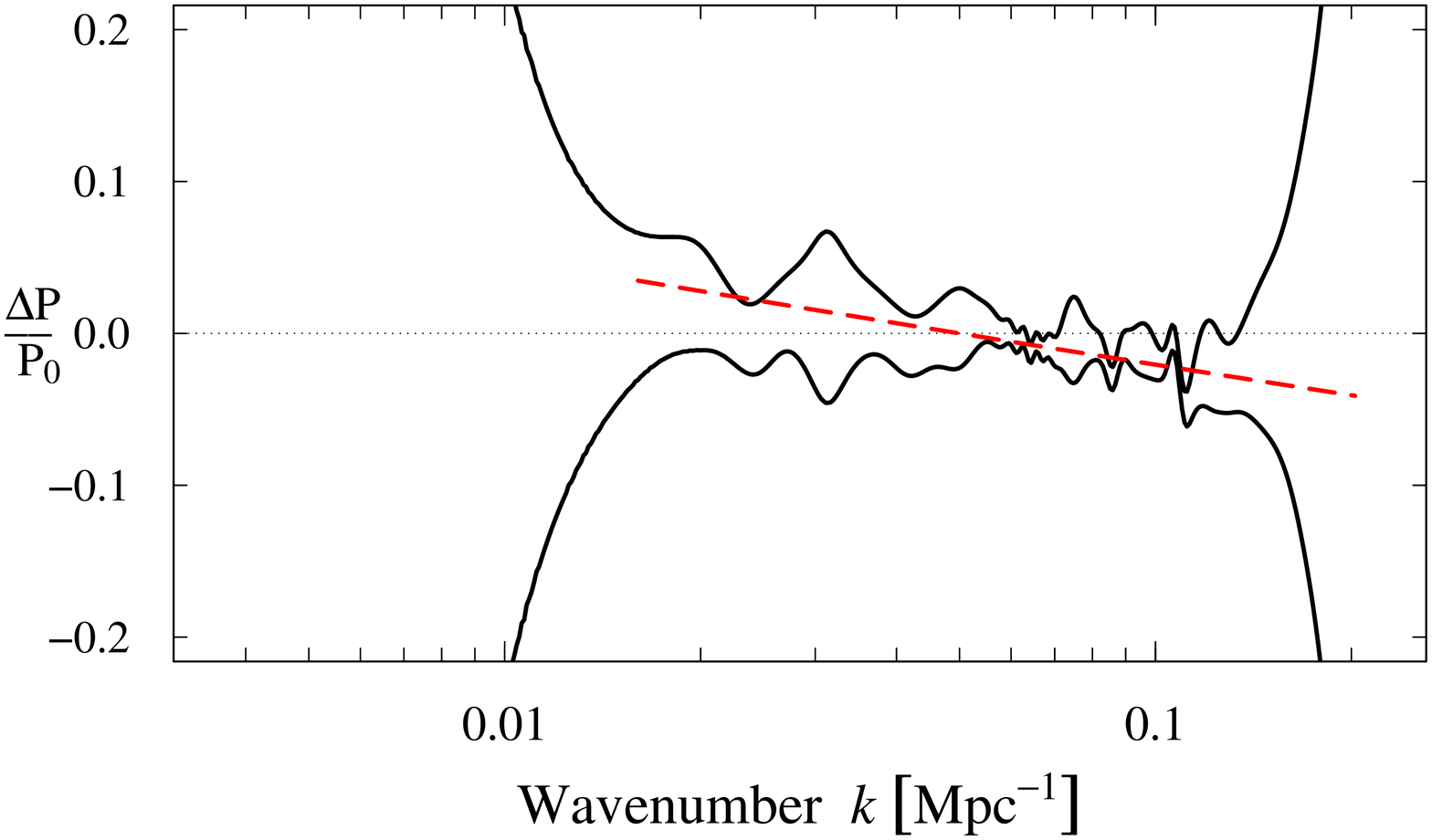}
\caption{Illustrating the estimated departures from scale-invariance in $k$-space
  for the case of an input $n_{{\rm S}}=0.97$
  spectrum (inclined dashed line), as in Fig.~\ref{fig:HZ8}.
  A tilt is recovered in the region $k=0.06-0.1$ Mpc$^{-1}$ with enough
  signal to noise to overrule the assumption of scale-invariance in model
  equation (\ref{eqn:PofkParametrisation}).
}\label{fig:ns0978}
\end{figure}

The $n_{{\rm S}}=0.97$ case is more delicate since we know in advance that
the power spectrum model equation (\ref{eqn:PofkParametrisation})
will not be able to accurately describe a tilted spectrum on large or small
scales. We can therefore expect some biasing in the recovery of the
cosmological parameters which will necessarily adjust to provide the
overall excess of power on large scales relative to small scales;
this is just the usual degeneracy between cosmological and power spectrum
parameters.

In fact to get reasonable results at all, we found it necessary to
apply equation~(\ref{eqn:degrade}) in order to orthogonalise the PCA modes to
the effect of the primordial power spectrum amplitude ${\mathcal
  P_0}$. The qualitative effect on the PCA modes is the the positive
definite mode 1 is removed. Having modified the PCA modes in this way,
the cosmological parameters are recovered
as $\omega_{\rm  B}h^2=0.0247\pm0.0002$, $\omega_{\rm
  D}h^2=0.116\pm0.001$, $H_0=74.6\pm0.7$, $\tau=0.183\pm0.006$,
${\mathcal P}/{\mathcal   P_0}=1.02\pm0.01$ for the case of using
eight PCA modes, showing biases at the 3$\sigma$ to 4$\sigma$ level.
The fact that the recovered dark matter
density shifts from $\Omega_{{\rm D}}h^2=0.113\pm0.001\rightarrow
0.116\pm0.001$ as the number of
PCA modes is increased provides a useful indication that there are
problems afoot with our power spectrum model
equation (\ref{eqn:PofkParametrisation}). 

Interestingly however, the PCA mode amplitudes are still very well
recovered, and we illustrate in Fig.~\ref{fig:ns097PCASpectrum8}
that the first ten mode amplitudes, if somewhat attenuated in
amplitude, provide strong evidence for a
power-law primordial power spectrum, showing a distinctive pattern
deviating from scale-invariance, $m_{{\rm a}}=0$. 
The corresponding departures from scale-invariance are shown in
Fig.~\ref{fig:ns0978} where the recovered power spectrum shows strong
evidence for a tilt, modulo some attenuation and oscillations in
regions of lower sensitivity. In short there is enough signal to noise to
overrule our assumption of scale-invariance, supplying us with strong evidence
that model of equation (\ref{eqn:PofkParametrisation}) needs
refining. It is likely that in a more refined analysis, one should
orthogonalise the PCA modes to the effect of the spectral index and
the other cosmological parameters in order to recover unbiased
estimates of the cosmological parameters.

\subsection{The Gaussian bump case}

\begin{figure}
\centering
\includegraphics[width=\linewidth]{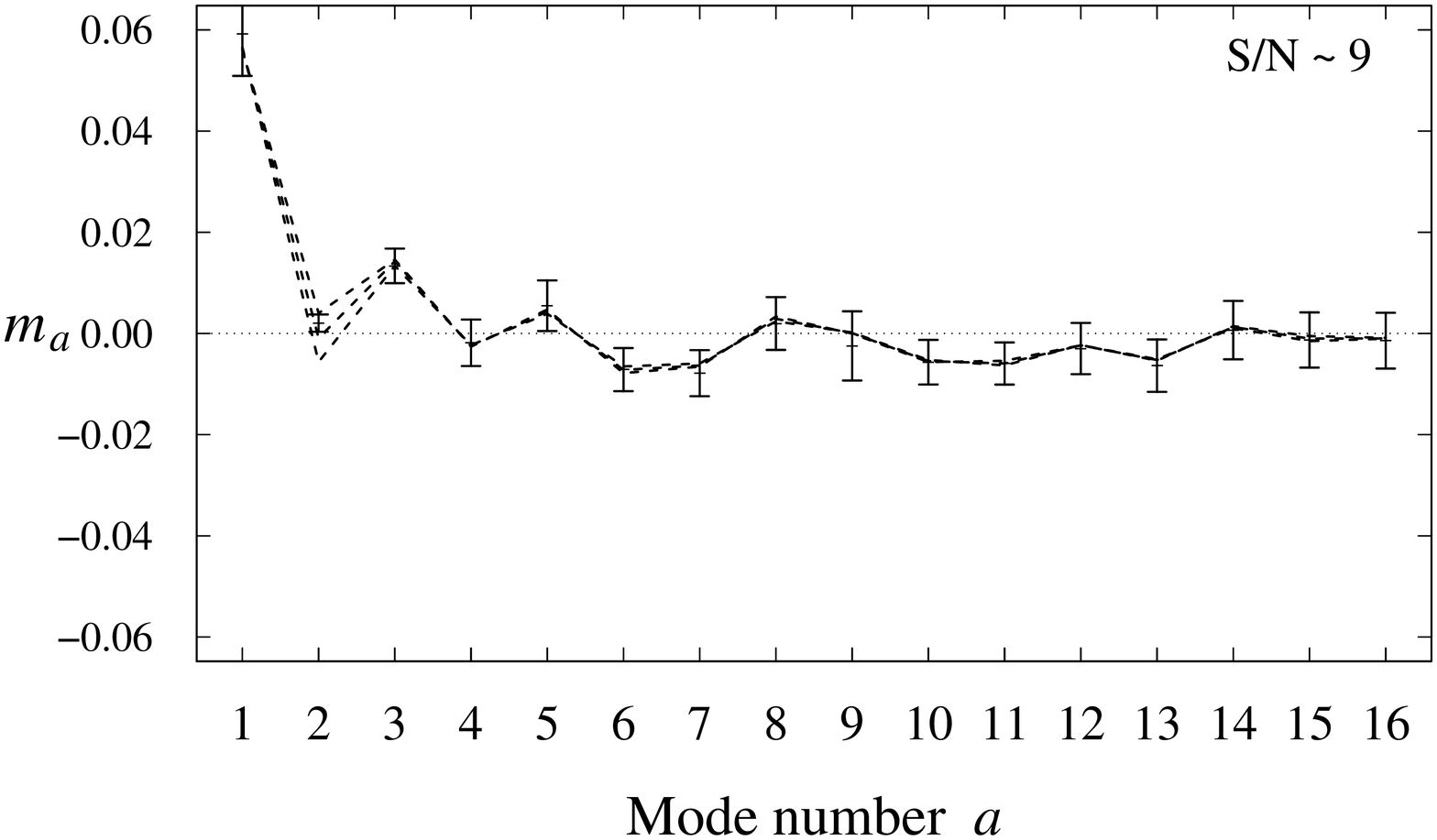}
\caption{Illustrating the recovery of the first sixteen principal
  component amplitudes from simulated \emph{Planck} data with an input
  Gaussian bump primordial power spectrum, as in
  Fig.~\ref{fig:PLANCK89_PCAspectrum}. The models  (dashed lines)
  correspond to the Gaussian bump of equation (\ref{eq:bump}),
  centred on $k_0 =\{0.082,0.08,0.078\}\,$Mpc$^{-1}$ (top to bottom, mode 2).
}\label{fig:bump12}
\end{figure}

\begin{figure}
\centering
\includegraphics[width=\linewidth]{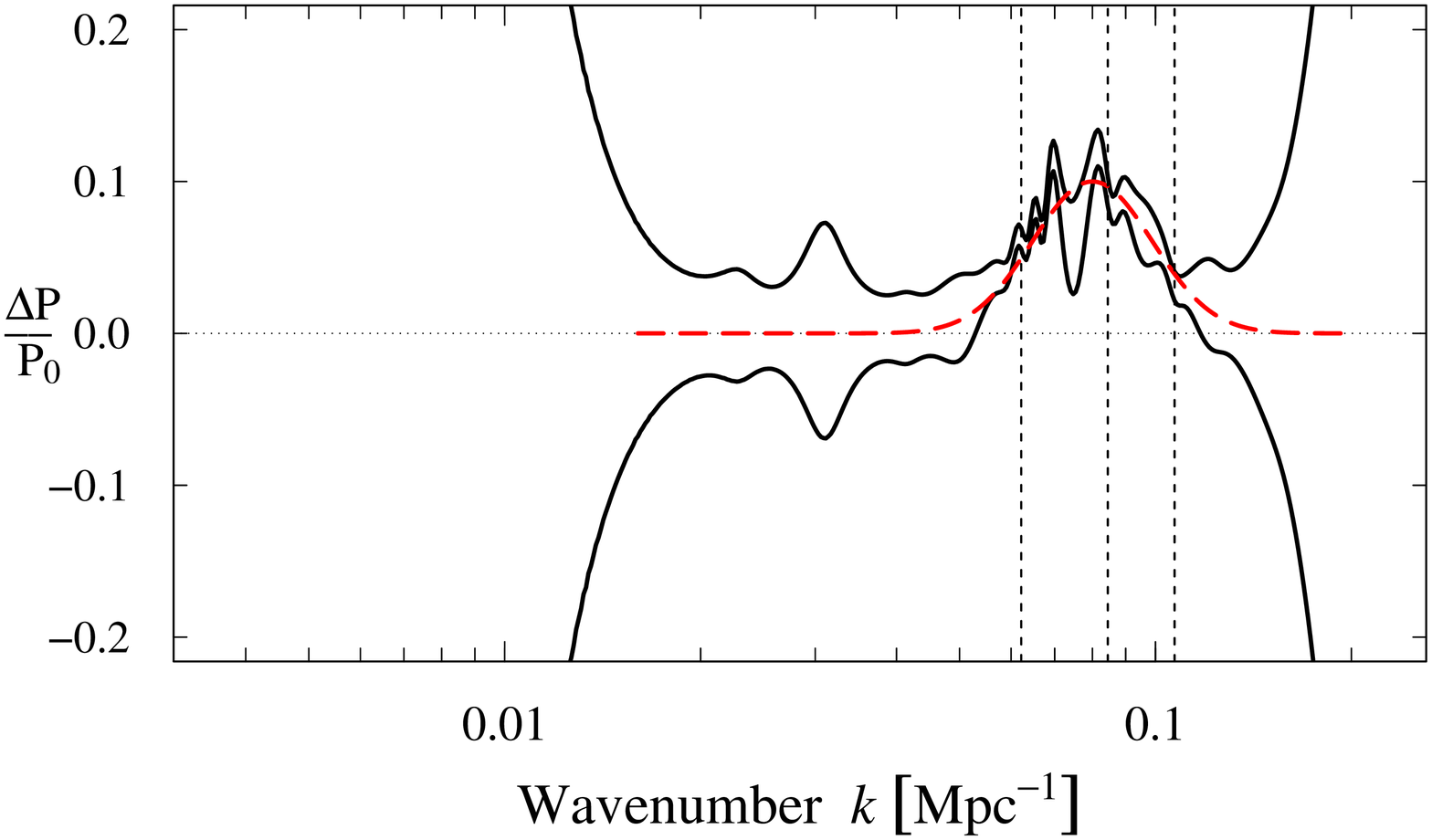}
\caption{Illustrating the estimated departures from scale-invariance in $k$-space
  for the case of an input scale-invariant plus Gaussian bump spectrum
  (dashed curve), as in Fig.~\ref{fig:HZ8}. A distinct bump like
  feature is recovered in the acoustic peak region. Precision
  polarization data would be required in order to better recover the feature in
  between the third, fourth, and fifth temperature acoustic peak scales
  (vertical dotted lines).
}\label{fig:bumprecon}
\end{figure}

Although completely contrived, this is perhaps the most interesting and
challenging case since the input primordial power spectrum now contains
distinct feature within the acoustic peak region. We illustrate in
Fig.~\ref{fig:bump12} that the first sixteen PCA amplitudes are
nonetheless rather well recovered and are consistent with the input Gaussian bump
model. In this case we can see that, for instance, the second PCA mode
strongly constrains the central position of the feature in $k$-space. In
Fig.~\ref{fig:bumprecon} we show that a bump like feature has indeed
been recovered, again modulo some attenuation and oscillations in
regions of lower sensitivity. The cosmological parameters are also very well
recovered with $\omega_{\rm B}h^2=0.0238\pm0.0002$, $\omega_{\rm
  D}h^2=0.122\pm0.002$, $H_0=71.6\pm0.9$, $\tau=0.170\pm0.005$,
${\mathcal P}/{\mathcal  P_0}=1.00\pm0.01$. This represents an
interesting success for the PCA method. 

\subsection{Summary and discussion}

To summarise the tests so far, the PCA method has been demonstrated
here to be very suitable and effective for measuring departures from
scale-invariance, both scale-free and scale-dependent, in the most
data-weighted regions of the $C_{\ell}$ spectrum.
In a realistic data analysis setup the recovered PCA mode amplitudes, 
together with the PCA modes themselves will represent an extremely powerful
compression of our information concerning the primordial power spectrum. 
At first sight this may represent an unnecessary data analysis stage
compared the usual parameter determination methods where one fits to the
$C_{\ell}$ data directly using the power spectrum model parameters on the
same footing as the other cosmological parameters. However, the point here
is to obtain first a detailed picture of the most important
departures from scale-invariance in the primordial power spectrum while at the
same time being able to weigh up the relative importance as well as
locating both in $k$ and $\ell$ space any
possible glitches or residual systematic effects in the $C_{\ell}$
data; then in the final data compression stage we can use the PCA
mode amplitudes to rapidly test any wide class of specific power spectrum models
with great ease and without recourse to any further $C_{\ell}$ likelihood
evaluations, as was recently emphasised by Kadota et~al (2005) for the
case of inflation models.

\section{Application to the \emph{WMAP} data}
 \label{sec:application}

In this Section we apply the PCA method to the currently available
temperature and temperature-polarization cross correlation spectra from
\emph{WMAP} (Kogut et~al 2003; Verde et~al 2003; Hinshaw et~al 2003)
and bandpowers in the range $600<\ell<2000$ from the VSA (Grainge et~al 2003;
Dickinson et~al 2004) ACBAR (Kuo et~al 2004), CBI (Pearson et~al
2003; Readhead et~al 2004) and Boomerang B2K (Jones et~al 2005,
Piacentini et~al 2005, Montroy et~al 2005) instruments.

To emphasise once more, we are working within the framework of spatially
flat $\Lambda$CDM cosmologies, described by five basic cosmological
parameters: the baryon density $\Omega_{\rm B}h^2$, the cold dark
matter density $\Omega_{\rm D}h^2$, the optical depth to last
scattering $\tau$, the ratio of the sound horizon to angular diameter
distance at last scattering $\theta=100r_{{\rm s}}^*/D_{{a}}^*$
(instead of $H_0$) and
the overall amplitude of scalar perturbations ${\mathcal P}_0$. In
addition we throw into the mix the first four PCA modes generated with
a noise model for \emph{WMAP} given by $\sigma_{{\rm
    noise}}^2=8.4\times10^{-3}(\mu$K-rad$)^2$ and $\theta = 13'$. 

\begin{figure}
\centering
\includegraphics[width=\linewidth]{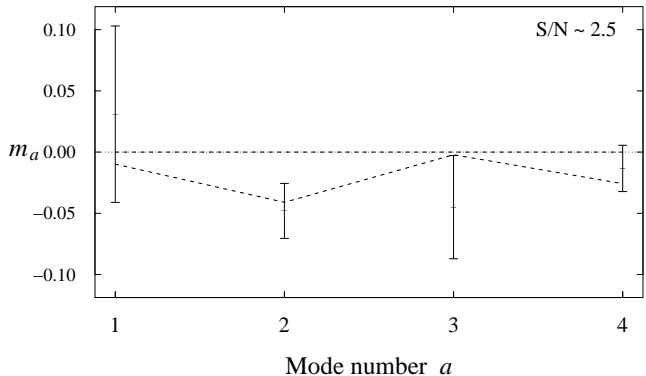}
\caption{Illustrating the current PCA measurements using current data from
  \emph{WMAP}, VSA, ACBAR, CBI and Boomerang. The compressed CMB data are well fit by $m_{\rm a}=0$
  and so show no evidence for breaking of scale-invariance. The dashed
  lines show power-law models with $n_{{\rm
      S}}(k_0=0.04$ Mpc$^{-1})=\{1.0, 0.94\}$ (top to bottom).
}\label{fig:WMAP4modes}
\end{figure}

\begin{figure}
\centering
\includegraphics[width=\linewidth]{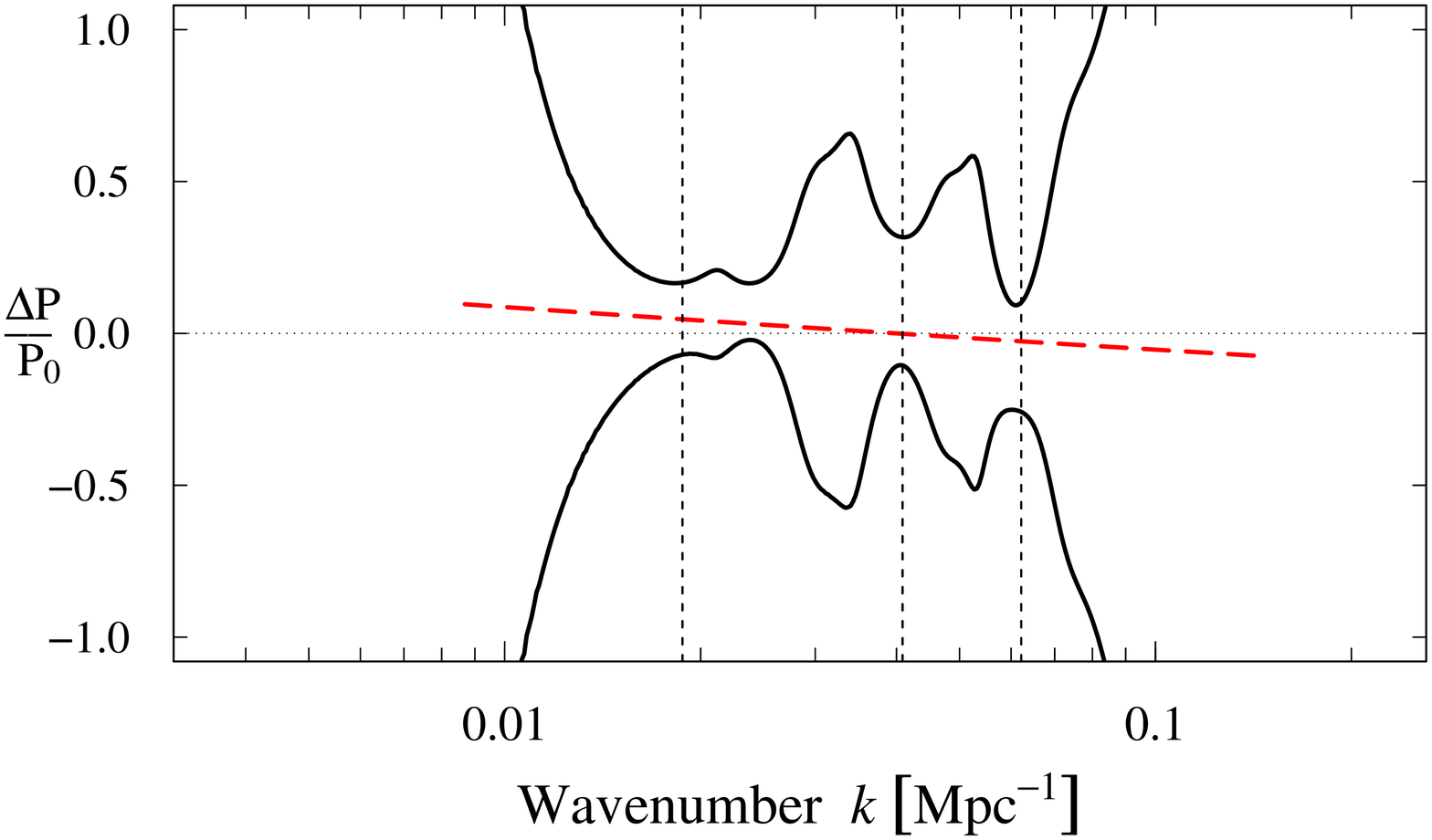}
\caption{Illustrating the estimated departures from scale-invariance
  using current data from \emph{WMAP}, VSA, ACBAR, CBI and Boomerang.
  The spectrum is scale-invariant showing only the slightest hint of a
  tilt. The best-fit spectrum with $n_{{\rm
      S}}(k_0=0.04$ Mpc$^{-1})=0.94\pm0.04$ is shown (dashed inclined
  line) as well as the first, second and third acoustic peak scales
  (vertical dotted lines).
}\label{fig:WMAPreconstruction}
\end{figure}

\begin{figure}
\centering
\includegraphics[width=\linewidth]{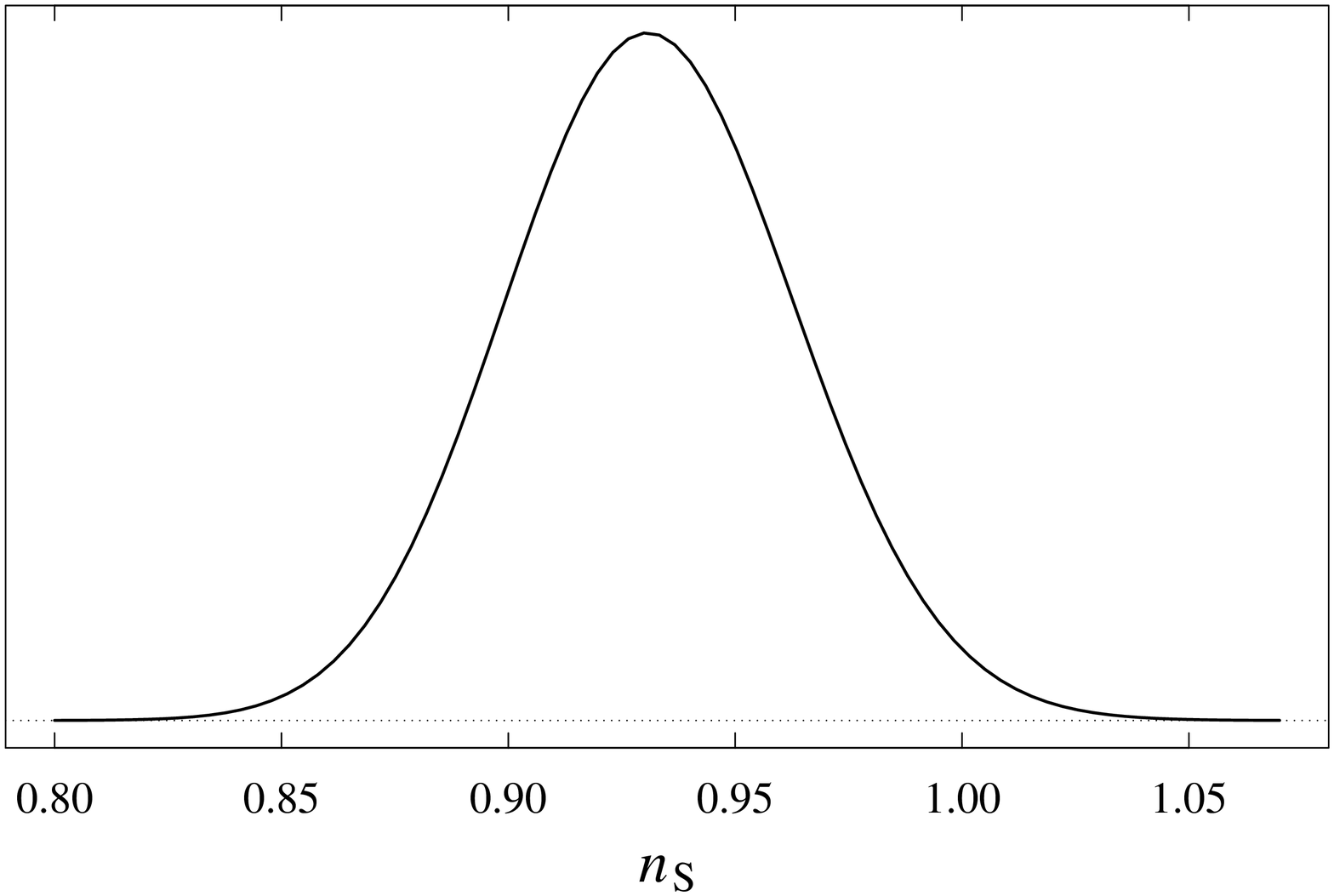}
\caption{Illustrating the posterior constraint on  the spectral
  index $n_{{\rm  S}}(k_0=0.04$ Mpc$^{-1})=0.94\pm0.04$ obtained from
  the four PCA mode amplitudes displayed in Fig.~\ref{fig:WMAP4modes}. 
}\label{fig:nslike}
\end{figure}

The measured amplitudes of the first four modes of Fig.~\ref{fig:modes1to4}
are displayed in Fig.~\ref{fig:WMAP4modes} with the corresponding
power spectrum in Fig.~\ref{fig:WMAPreconstruction}. The broad
picture painted here is 
that we find no evidence for the breaking of scale-invariance: the
mode amplitudes are very well fit my $m_a=0$. Only a single mode on scales
corresponding to the second acoustic peak shows an S/N $>1$, which is
barely worth mentioning aside from the fact that 
it can easily be accommodated by a slightly red primordial power
spectrum: projecting power-law primordial power spectra onto the PCA
basis and using a simple Gaussian likelihood function we find the constraint
on the spectral index to be $n_{{\rm
    S}}(k_0=0.04$ Mpc$^{-1})=0.94\pm0.04$, displayed in
Fig.~\ref{fig:nslike}, and which is in accordance with
conventional studies of the primordial power spectrum. It is
also possible to make a detailed comparison with the primordial power
spectrum bandpowers from fig.~4 of Bridle et~al
(2003), as well as with orthogonal wavelet expansion constraints in
fig.~2 of Mukherjee \& Wang (2003b). We all find the same very weak trend for a
20-30$\%$ drop in power between between the first acoustic peak at
$k=0.02$ Mpc$^{-1}$ and the third acoustic peak scale at
$k=0.07$ Mpc$^{-1}$. Again, the trend is not so much interesting at this
stage as the consistency between these complementary methods.

\section{Conclusions}
 \label{sec:conclusions}
In this work we have implemented and investigated a principle
component analysis (PCA) technique in order to study the possible departures
from scale-invariance that may exist in the spectrum of primordial
curvature perturbations, which are observable via the CMB
anisotropies. The essence of this method is to decompose the primordial
power spectrum into a scale-invariant component plus a series of
orthonormal modes which reflect our expectation of where the
departures from scale-invariance are likely to be best probed by the
data. The information from the CMB is then be compressed into a series
of mode amplitudes which can easily be compared with predictions from
any wide class of primordial power spectra without recourse to any
further $C_{\ell}$ likelihood evaluations.  

The method was first tested on simulated \emph{Planck} data using an input
scale-invariant spectrum and we observed good performance in the
simultaneous recovery of cosmological parameters and the principal component mode
amplitudes via an MCMC exploration of the full parameter space.
In the case of simulated data from an input power-law spectrum with
spectral index $n_{{\rm S}}=0.97$, the recovery of the cosmological
parameters was biased as they adjusted to provide an overall excess of
large-scale to small-scale power. However, the biasing is evidenced by
fluctuating cosmological parameter constraints as the number of power spectrum
principal components is increased. Moreover, the PCA mode amplitudes
were still very well recovered, showing strong evidence for a tilted
primordial power spectrum and providing enough signal to noise to
overrule our assumption of scale-invariance.
Thus PCA can be used as a self-consistent means for justifying a more
refined power spectrum model than the one considered here in
equation (\ref{eqn:PofkParametrisation}). We also demonstrated that the PCA
method is capable of measuring departures from scale-free spectra by
considering simulated data from a primordial power spectrum containing
a $10\%$ gaussian bump in the acoustic peak region, and observing good
recovery of both the PCA mode amplitudes and the cosmological parameters.

Finally, as a proof of concept of the method we provided a first
glimpse of the principal component mode amplitudes that can be
obtained from the currently available CMB data from \emph{WMAP}, VSA,
ACBAR, CBI and Boomerang. We obtained measurements of the first four principal
components corresponding to scales across the first and second
acoustic peaks, finding no evidence for the breaking of
scale-invariance with only a hint of a red primordial power spectrum with 
spectral index $n_{{\rm  S}}(k_0=0.04 $ Mpc$^{-1})=0.94\pm0.04$,
consistent with other studies in the literature, with a
total signal to noise at not more than S/N $\sim2.5$.

Assuming that the Gaussian adiabatic density perturbation scenario
continues to hold as our observations of the CMB improve in the near
future, then we will soon move into the regime where the information
about the primordial power spectrum will completely outweigh the
information about the cosmological parameters which become, from this
perspective, well-understood nuisance parameters to be carefully
integrated out. It seems very likely that principal component
analysis, or else another very similar data compression technique, will
be essential for fully exploiting the forthcoming temperature and
polarization $C_{\ell}$ data.

\section*{Acknowledgements}
I thank Carlo Baccigalupi, Sergei Bashinsky, Uro\v{s} \mbox{Seljak}, Roberto
Trotta and Ben Wandelt for their helpful \mbox{discussions} and comments,
Antony Lewis for supplying his
\emph{Planck} simulations code via {\tt www.cosmocoffee.org}, Christophe
Ringeval for his help with the modifications to accuracy of the {\sc
  CAMB} code, and the University of Geneva, where this work was
started,  and the Trieste Observatory
for hosting the main computations in this work. I acknowledge the use of the
Legacy Archive for  Background Data Analysis (LAMBDA). Support for
LAMBDA is provided by the Office of Space Science. I also acknowledge
the use of the {\sc R} statistical computing environment in which much of
this analysis was  performed, and the {\sc CosmoMC} (Lewis \& Bridle
2002) and {\sc CAMB} (Lewis, Challinor \& Lasenby 2000) packages.

{}

\end{document}